# Ising and regular solution models revisited for 2D open systems

Andriy Gusak[1,2,*], Serhii Abakumov[2]

[1.] Ensemble3 Centre of Excellence, Wolczynska Str. 133, 01-919 Warsaw, Poland

[2.] Bohdan Khmelnytsky National University of Cherkasy, 81 Shevchenko Blvd., UA-18031 Cherkasy, Ukraine

*-corresponding author

**Abstract**

The Ising model is well-known for illustrating the fundamental characteristics of phase transitions in closed systems. In this article, we propose a generalization of the 2D Ising model to open systems, considering the divergence of in- and out-fluxes within a mean-field approximation, and explore its application for classifying steady states based on external flux (deposition rate), temperature, and composition. We focus on cases with positive mixing energy, which lead to the typical spinodal and binodal domes at the phase diagram under the regular solution approximation for closed systems. For open systems, we demonstrate that a supercritical external flux divergence stabilizes the system, preventing decomposition. We identify rate-dependent spinodal and binodal domes, as well as a subdivision of the instability region at subcritical rates into three distinct steady-state morphologies: spots (gepard-like), layers (zebra-like) – labyrinth or lamellae, and mixed patterns (a combination of "gepard" and "zebra"). This morphology map depends on the initial conditions, demonstrating memory effects and hysteresis.

## 1. Introduction

To begin, let us define the terms 'closed' and 'open' systems as used in this paper. We refer to a thermodynamic system as 'closed' when it is subject to homogeneous external conditions across its boundaries. Typically, this includes cases such as: (a) an isolated system, (b) a system with a fixed volume within a thermal bath at a uniform temperature T, or (c) a system under fixed homogeneous pressure within a thermal bath at uniform temperature T. In case (a), the system evolves toward a state of maximum entropy; in case (b), toward minimum Helmholtz free energy, F=U−TS; and in case (c), toward minimum Gibbs free energy, G=U−TS+pV. By contrast, we describe a system as 'open' if it experiences inflow and outflow of matter and/or energy, typically driven by gradients in electrochemical potential, temperature, or stress. Such gradients cannot vanish over time due to boundary conditions, such as hot and cold ends of a system or opposite poles in an electrical circuit. While closed systems always progress toward equilibrium, open systems may reach a steady state, exhibit oscillatory behavior, or even fail, but they do not reach equilibrium. Thermodynamics, phase transitions, driving forces, partition functions, probability distributions, and the final equilibrium state of closed systems are well understood, although the principle governing the choice of evolutionary paths among multiple possibilities remains an open question. Many of the general behaviors of closed systems can be effectively studied using simple, fundamental models, such as the Ising model. However, for open systems, much less is known about the selection and evolution of final states. To gain insight into the general behaviors of open systems, we need a comparable foundational model – one as elementary as the Ising model. The Ising model serves as a classic example for illustrating essential thermodynamic and kinetic features of phase transitions in closed systems [1]. In this article, we introduce an 'Ising-like' model, maintaining similar simplifications, to study the principal features of open systems, particularly their steady states. For this purpose, we adopt Georges Martin's modified approach, based on master equations for site occupancy probabilities. Martin first proposed a self-consistent nonlinear kinetic model for a quasi-1D system in 1990 [2], which was later applied to thin-film nonlinear interdiffusion across contact zones with pronounced diffusion asymmetry, typically confined to a few atomic layers (Erdelyi, Beke et al. [3,4]). Later we developed a 3D model [5-8] which has since served as the basis for new software, SKMF (Stochastic Kinetic Mean-Field, skmf.eu), for atomistic simulations of diffusion-controlled transformations, including spinodal decomposition, nucleation, ripening, reactive diffusion, and phase competition. Recently, we applied the 2D version of SKMF to examine pattern formation during vapor co-deposition of binary alloys [9]. Pattern formation on the mesoscopic scale during crystallization from liquid or vapor phases has become a well-studied subject [10-15], yet atomic-scale pattern formation remains relatively underexplored [16]. In our previous work [9], we used a somewhat artificial model, assuming that a new atomic (001) plane in the FCC lattice is rapidly filled by incoming atoms, after which atomic exchanges occur over a time $\frac{\delta}{V}$ (where V is the deposition velocity, equal to the product of deposition flux density and the atomic volume of the solid phase, and δ is the interplanar spacing in the deposition direction). After this period, diffusion within the 'buried' atomic plane is considered to be fully frozen.

2. **Basic model assumptions**

In this version of our model, we aim to create a time-continuous framework, avoiding any stepwise kinetics involving sharp avalanches of atomic plane filling followed by isolated diffusion within the plane. Instead, we introduce a 'smeared' time scale where diffusion within the top surface layer and deposition occur simultaneously. This approach allows us to use a moving reference frame that travels with the top surface at a constant velocity V in the deposition direction. In this moving frame, in addition to diffusion fluxes along the plane (leading to partial decomposition), there are two external fluxes perpendicular to the top surface: an inflow $\frac{V}{\Omega}C^{dep}$ (where Ω is the atomic volume in the solid phase) and an outflow $\frac{V}{\Omega}C_A(i,j,k=0)$, with k=0 indicating the top plane and i,j denoting specific lattice sites within it (in an FCC lattice, i+j+k is even). Both processes—the diffusion along the top plane and the flux divergence across it—are mathematically represented by two terms on the right-hand side of Eq. (1), which governs the site occupancy probabilities:

$$\frac{\partial C_A}{\partial t} = \sum_{in=1}^{Z\|} \{-C_A(i)C_B(in)\Gamma_{AB}(A(i) \leftrightarrow B(in)) + C_B(i)C_A(in)\Gamma_{AB}(A(in) \leftrightarrow B(i))\} +$$

$$+ \frac{V}{\delta}(C^{dep} - C_A(i)) \quad (1)$$

Here $C_A(i)$ actually represents $C_A(i, j, k = 0)$ and is the probability of site (i,j) ($x = \frac{a}{2} * i, y = \frac{a}{2} * j$) in the top plane (k=0) ($z = \frac{a}{2} * 0$) being occupied by species A. We work here only with an FCC monocrystalline lattice grown by deposition in the <001> direction, and i+j+k is an even number. In our model, atomic exchanges happen only between sites in the same top plane (kn=0, in+jn+kn – even). Exchange frequencies are determined by a Boltzmann-like expression,

$$\Gamma_{AB}(A(i) \leftrightarrow B(in)) = v_0 exp(-\frac{E^S - (E_A(i) + E_B(in))}{kT}) \quad (2)$$

where $(E_A(i) + E_B(in)) = (E_A(i, j, k = 0) + E_B(in, jn, kn = 0))$ is the interaction energy of neighboring exchanging atoms before the exchange, and $E^S$ is a saddle-point energy during this exchange (assumed to be the same in the original Martin model and all its developments).

Energies are calculated in a mean-field approximation and contain $Z\| = 4$ nearest neighbors from the same top plane (k=0) with indexes $(i \pm 1, j)$ and $(i, j \pm 1)$, and also $Z\perp = 4$ other nearest neighbors from the plane below ($k = 1, z = -\frac{a}{2} * 1$) with indexes $(i \pm 1, j)$ and $(i, j \pm 1)$.

$$E_A(i) = \sum_{i'=1}^{Z\|+Z\perp} (C_A(i')V_{AA} + C_B(i')V_{AB}) =$$

$$= (Z\| + Z\perp)V_{AB} + (V_{AA} - V_{AB})\sum_{i'=1}^{Z\|+Z\perp} C_A(i') \quad (3)$$

$$E_B(in) = \sum_{in'=1}^{Z\|+Z\perp} (C_A(in')V_{BA} + C_B(in')V_{BB}) =$$

$$= (Z\| + Z\perp)V_{BB} + (V_{AB} - V_{AA})\sum_{in'=1}^{Z\|+Z\perp} C_A(in') \quad (4)$$

In the case of co-deposition of the (001) planes of an FCC lattice, as mentioned above, the number of nearest neighbors within the top plane (simultaneously the number of possible atomic exchanges) is $Z\| = 4$, and the number of nearest interacting neighbors in the preceding (subsurface) plane is $Z\perp = 4$. "i" and "in" are two neighboring sites within the top atomic plane, exchanging atoms. At a fixed "i" there are $Z\| = 4$ possibilities for "in". "i'" represents the nearest interacting neighbors of site "i", and their number is $Z = Z\| + Z\perp = 8$, similarly, "in'" represents the nearest interacting neighbors of site "in", and their number is also $Z = Z\| + Z\perp = 8$.

We seek the simplest possible model, avoiding extra calculations. To achieve self-consistency and limit unknown probabilities to only one atomic plane, we postulate (in our model) that the probabilities in the subsurface plane k=1 are entirely determined by their nearest neighbors in the top plane k=0:

$$C_A(i, j, k = 1) = \frac{1}{4} * (C_A(i + 1, j, k = 0) + C_A(i - 1, j, k = 0) +$$

$$+C_A(i, j + 1, k = 0) + C_A(i, j - 1, k = 0)) \quad (5)$$

This postulate is not absolute — it can be modified, for example, if we want to describe ordering, but in our current case of decompositions, it works well, as we will see below.

For simplicity, we also assume: $V_{AA}=0$, $V_{BB}=0$, $V_{AB}=E_{mix}$. Then

$$\Gamma_{AB}(A(i) \leftrightarrow B(in)) = v_0 e^{-E^S/kT} exp[\frac{E_{mix}}{kT}(Z - \sum_{i'=1}^{Z} C_A(i') + \sum_{in'=1}^{Z} C_A(in'))],$$

$$\Gamma_{AB}(A(in) \leftrightarrow B(i)) = v_0 e^{-E^S/kT} exp[\frac{E_{mix}}{kT}(Z - \sum_{in'=1}^{Z} C_A(in') + \sum_{i'=1}^{Z} C_A(i'))] \quad (6)$$

Then, the master equation for occupancy probabilities within the surface layer k=0 is:

$$\frac{\partial C_A(i)}{\partial tt} = \sum_{in=1}^{Z\parallel} \{-C_A(i)(1 - C_A(in)) \cdot exp[\frac{E_{mix}}{kT}(\sum_{in'=1}^{Z} C_A(in') - \sum_{i'=1}^{Z} C_A(i'))] +$$

$$+(1 - C_A(i))C_A(in) exp[\frac{E_{mix}}{kT}(\sum_{i'=1}^{Z} C_A(i') - \sum_{in'=1}^{Z} C_A(in'))]\} + v * (C^{dep} - C_A(i)) \quad (7)$$

Here, the non-dimensional time and non-dimensional velocity parameters are:

$$tt = t \cdot v_0 exp[\frac{ZE_{mix}-E^S}{kT}], v = \frac{V}{v_0 \delta exp[\frac{ZE_{mix}-E^S}{kT}]} \quad (8)$$

In the present article, we will consider only the case of positive mixing energy, corresponding to a tendency towards decomposition, which can be realized partially or not realized depending on temperature, composition, and deposition velocity.

### 3. Binodal and spinodal in KMF model at zero rate (decomposition in closed system)

We start by checking if our model correctly describes the classic case of a closed system (V=0), which should correspond to the standard regular solution approximation — decomposition cupola, including the spinodal (absolute instability criterion) and binodal (equilibrium criterion).

### 3.1. Binodal (thermodynamic equilibrium which can be reformulated as detailed flux balance).

First of all, let us consider the equilibrium condition at V=0. It corresponds to balance equations (following from eq. (1)):

$$C_A(i)(1 - C_A(in))\Gamma_{AB}(A(i) \leftrightarrow B(in)) = (1 - C_A(i))C_A(in)\Gamma_{AB}(A(in) \leftrightarrow B(i)) \quad (9)$$

or $C_A(i)(1 - C_A(in))\exp\left(-\frac{E^s - (E_A(i) + E_B(in))}{kT}\right) = (1 - C_A(i))C_A(in)\exp\left(-\frac{E^s - (E_A(in) + E_B(i))}{kT}\right)$

which can be reformulated as

$$\left(E_A(i) + kT\ln C_A(i)\right) - \left(E_B(i) + kT\ln C_B(i)\right) =$$
$$= \left(E_A(in) + kT\ln C_A(in)\right) - \left(E_B(in) + kT\ln C_B(in)\right). \quad (10)$$

Since $E_A(i) + kT\ln C_A(i) = \mu_A(i)$ is a local chemical potential of A and $E_B(i) + kT\ln C_B(i) = \mu_B(i)$ – that of B, their difference $\mu_{AB}(i) \equiv \mu_A(i) - \mu_B(i)$ is just a reduced chemical potential (change of Gibbs free energy due to the replacement of atom B by atom A). Equality of reduced chemical potential in different sites means thermodynamic equilibrium, including the case when these sites belong to two different phases (solid solutions at two different wings of the binodal). In this case, $C_A(i) = C_A(i') = C_A(\alpha), C_A(in) = C_A(in') = C_A(\beta) = 1 - C_A(\alpha)$. Substituting it into eq. (7) under conditions $v = 0, \frac{\partial C_A(i)}{\partial tt} = 0$, after simple algebra one gets

$$\frac{C_A(\alpha)}{1 - C_A(\alpha)} = \exp\left[-\frac{8E_{mix}}{kT}(1 - 2C_A(\alpha))\right], \quad (11)$$

and this expression coincides with the binodal equation for the regular solution model with 8 nearest neighbors for each site.

### 3.2. Instability criterion for closed system (spinodal)

Usually, people prescribe an idea of infinitesimal concentration waves, which may exponentially increase or decrease, depending on the wave vector, temperature, and composition, to Cahn and Hilliard in their phenomenological analysis of spinodal decomposition [17]. Actually, an analogous idea of the instability criterion for the non-linear kinetic equations was suggested much earlier for the atomic scale by Anatoliy Vlasov in his non-local statistical approach to crystals [18,19]. Later, analogous ideas were also used at the atomic scale by Armen Khachaturyan in his concept of concentration waves [20]. We will be looking for a solution in the form of an atomic-scale concentration wave with an infinitesimal time-dependent amplitude A:

$$C_A(i, j, k = 0) = C^{dep} + \delta C(i, j, k = 0) = C^{dep} + A(tt, \vec{q})\exp(I\vec{q}\vec{r_{i,j}}) =$$
$$= C^{dep} + A(tt, q_x, q_y)\exp\left(I\frac{a}{2}(q_x i + q_y j)\right). \quad (12)$$

Here, I is an imaginary number, sqrt(-1). Then, according to our condition (5), for the sites of sublevel (k=1), the concentration wave is as follows:

$$C_A(i',j',k=1) = \frac{1}{4} * \Big(C_A(i'+1,j',k=0) + C_A(i'-1,j',k=0) + C_A(i',j'+1,k=0) +$$

$$C_A(i',j'-1,k=0) = C^{dep} + A\exp\left(I\frac{a}{2}(q_x i' + q_y j')\right) * \frac{1}{4} * \left(\exp\left(I\frac{a}{2}q_x\right) + \exp\left(-I\frac{a}{2}q_x\right) + \exp\left(I\frac{a}{2}q_y\right) + \exp\left(-I\frac{a}{2}q_y\right)\right) =$$

$$= C^{dep} + A\exp\left(I\frac{a}{2}(q_x i' + q_y j')\right) * \frac{1}{2} * \left(\cos\left(\frac{a}{2}q_x\right) + \cos\left(\frac{a}{2}q_y\right)\right). \tag{13}$$

We substitute eqs. (12) and (13) into eq. (7), expanding everything into series over small (A) and neglecting the second-order and higher-order terms. Simple algebra leads to the following stability/instability criterion for the amplitudes of concentration fluctuation waves:

$$\frac{\partial \ln A}{\partial tt} = 4\left[1 - \cos\left(q_x \frac{a}{2}\right)\cos\left(q_y \frac{a}{2}\right)\right] *$$

$$* \left\{\frac{8E_{mix}}{kT} C^{dep}(1 - C^{dep})\left(\cos\left(q_x \frac{a}{2}\right)\cos\left(q_y \frac{a}{2}\right) + \frac{1}{4}(\cos\left(q_x \frac{a}{2}\right) + \cos\left(q_y \frac{a}{2}\right))^2\right) - 1\right\} \tag{14}$$

Instability case: $\frac{\partial \ln A}{\partial tt} > 0$. In the long-wave approximation

$(qa \ll \pi, \cos\left(q_x \frac{a}{2}\right) \approx 1 - (q_x \frac{a}{2})^2/2)$ it means $\frac{16 E_{mix}}{kT} C^{dep}(1 - C^{dep}) > 1$, which **coincides with the spinodal criterion in the regular solid solution model at Z=8**.

4. **Rate-dependent binodal in open system.**

Here we discuss a binodal-like solution – a steady state instead of equilibrium. As follows from eq. (7), in the case of open systems, the "binodal-like" condition contains an additional rate-dependent term:

$$v * (C_A(i) - C^{dep}) = \sum_{in=1}^{Z\|} \{-C_A(i)(1 - C_A(in)) \cdot \exp\left[\frac{E_{mix}}{kT}\left(\sum_{in'=1}^{Z} C_A(in') - \sum_{i'=1}^{Z} C_A(i')\right)\right] +$$

$$+ (1 - C_A(i))C_A(in) \exp\left[\frac{E_{mix}}{kT}\left(\sum_{i'=1}^{Z} C_A(i') - \sum_{in'=1}^{Z} C_A(in')\right)\right]\} \quad . \tag{15}$$

So far, we cannot suggest as natural interpretation of this condition as was the detailed balance and equality of reduced chemical potentials for the closed system in eqs. (9) and (10). So, we just simulated decomposition numerically by solving the set of equations (7) and tracking the solution until it practically satisfied eq. (15). In that process, we fixed the maximal (right part of binodal) and minimal (left part of binodal) concentrations within our system (with small corrections due to noise and to Gibbs-Tomson corrections at the curved interfaces). Of course, tending to the steady state of the solution of eq. (9) should lead to the validity of eq. (15), at least in the case when initial inhomogeneities are large enough to overcome the nucleation barrier of decomposition. The result – the rate-dependent binodals (as well as rate-dependent spinodals in section 5) is shown in Fig. 1. We emphasize an important difference between binodal in closed and open systems: In closed system any point below binodal corresponds to alloy which will decompose - via spinodal decomposition of absolutely unstable solution, if this point is simultaneously below spinodal, or via nucleation, precipitation and growth in the metastable solution, if this point is between binodal and spinodal. In both cases the result is same – system transforms into two-phase state with two marginal compositions corresponding to binodal. In open system, as we will see in Section 7 (Fig.5), the situation is ambiguous: part of interdomes region (beyond spinodal but within binodal) indeed leads to decomposition (if one uses the preexisting structures as the initial condition), and another part demonstrates the full absence of decomposition. In this sense, the **rate-dependent** binodal is "smeared", like the **size-dependent** binodal for nanoparticles [21-23]

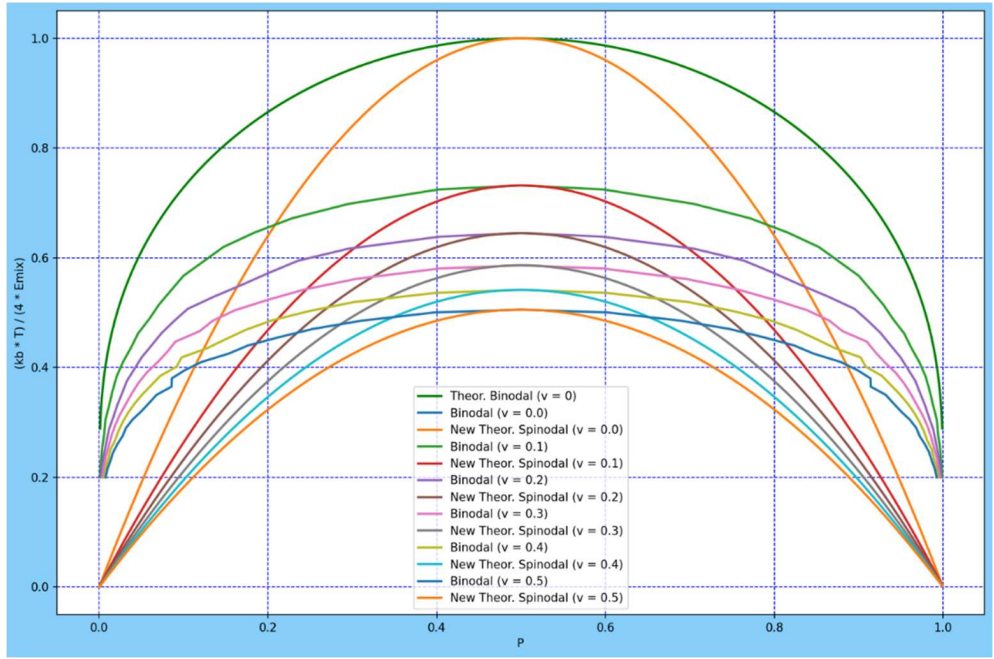

*Fig. 1. Rate-dependent binodals and rate-dependent spinodals at $v = 0, 0.1, 0.2, 0.3, 0.4, 0.5, ...$*

## 5. Rate-dependent spinodal - Instability in respect to infinitesimal perturbations

In full analogy with subsection 3.2, for the open system, the criterion of absolute instability and the following decomposition is reduced to the positive sign of the following derivative:

$$\frac{\partial lnA}{\partial tt} = -v + 4\left[1 - \cos\left(q_x \frac{a}{2}\right)\cos\left(q_y \frac{a}{2}\right)\right] *$$

$$* \left\{\frac{8E_{mix}}{kT} C^{dep}(1 - C^{dep})\left(\cos\left(q_x \frac{a}{2}\right)\cos\left(q_y \frac{a}{2}\right) + \frac{1}{4}(\cos\left(q_x \frac{a}{2}\right) + \cos\left(q_y \frac{a}{2}\right))^2\right) - 1\right\} \quad (16)$$

It means that at a fixed composition of deposition flux, and at a fixed temperature, the alloy can be stabilized against decomposition by the velocity

$$v > v^{critical} = \max(in\ respect\ to\ q_x, q_y) \left\{ 4\left[1 - \cos\left(q_x \frac{a}{2}\right)\cos\left(q_y \frac{a}{2}\right)\right] * \left[\frac{8E_{mix}}{kT} C^{dep}(1 - C^{dep})\left(\cos\left(q_x \frac{a}{2}\right)\cos\left(q_y \frac{a}{2}\right) + \frac{1}{4}(\cos\left(q_x \frac{a}{2}\right) + \cos\left(q_y \frac{a}{2}\right))^2\right) - 1\right]\right\} \quad (17)$$

One may show that for the concentration waves along the diagonal direction <110>, $\vec{q} = \left(\frac{q}{\sqrt{2}}, \frac{q}{\sqrt{2}}, 0\right)$, the maximal growth rate occurs at

$$\cos^2\left(\frac{qa}{2\sqrt{2}}\right) = \frac{W+1}{2W}, \text{ (here W}=\frac{16E_{mix}}{kT} C^{dep}(1 - C^{dep}) > 1)$$

At that, the minimal critical velocity is

$$v^{critical}(<110>) = \frac{(W-1)^2}{W} \quad (18)$$

On the other hand, for the concentration waves along the direction <100>, $\vec{q} \equiv (q, 0, 0)$, have the maximal growth rate occurs at $\cos\left(\frac{qa}{2}\right) = \frac{\sqrt{40+2/W}-5}{3}$. At that, the minimal critical velocity is

$$v^{critical}(<100>) = \frac{16}{27} W\left(4 - \sqrt{10 + \frac{6}{W}}\right)\left(\sqrt{10 + \frac{6}{W}} - 1 - \frac{3}{W}\right) \quad (19)$$

Comparison of eqs. (18) and (19) in Fig. 2 shows that the direction <100> should win the competition with <110>:

$$v^{critical}(<100>) > v^{critical}(<110>)$$

Indeed, among various morphologies, we often see the lamellae along X or Y but never along the diagonals – see below. (In another version of our model, containing only 4 nearest neighbors within the upper plane for interactions as well as for exchanges, critical velocities appear to be the same for both directions, so that the lamellae along the diagonals as well as along X or Y should be observed. – to be described elsewhere)

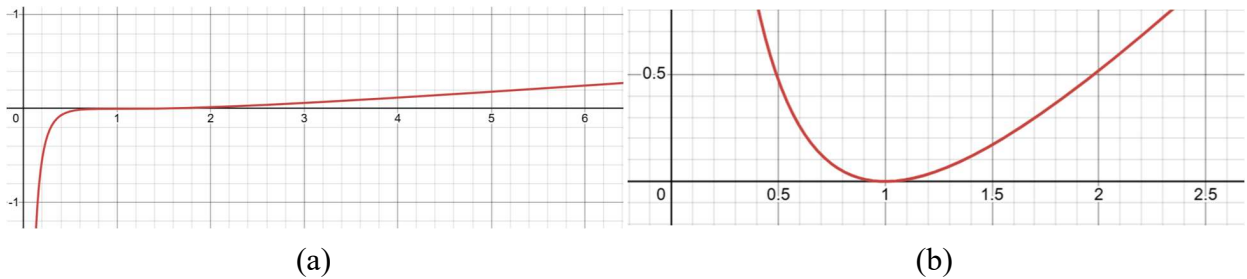

(a)          (b)

*Fig. 2. (a) Difference between the critical velocities for concentration wave amplification in the directions <100> (eq. (19)) and <110> (eq. (18)) as a function of $W = \frac{16E_{mix}}{kT} C^{dep}(1 - C^{dep})$, $(W > 1)$ (b) dependence of $v^{critical}(<100>)$ on the value of W (eq. (19)).*

From eq. (19), we obtain the velocity-dependent spinodal curve for the open system:

$$\frac{kT}{4Emix} \approx \frac{4C(1-C)}{W(v)}, \text{ where } W(v) \text{ is a solution of the transcendental eq. (19)} \quad (20)$$

We checked that at least for $0 < v < 0.5$ the difference with the <110> case, which corresponds to the analytic expression

$$\frac{kT}{4Emi} \approx \frac{4C(1-C)}{1+\frac{v}{2}+\left(\sqrt{(1+\frac{v}{2})^2-1}\right)}, \quad (21)$$

is less than 1 percent. Therefore, for $v \leq 0.5$ one may use this analytic approximation.

## 6. Morphology maps for steady-state. (Amplitude of initial noise 0.001, dynamic noise zero.)

If we use a homogeneous alloy with small initial fluctuations of composition (say, 0.01), the decomposition in steady-state is found only for concentrations and temperatures under the rate-dependent spinodals. These regions are divided into three types of final steady morphologies: Gepard (spots), mixed morphology Gepard+Zebra, and Zebra (layered morphology – labyrinth or lamellae) – Figs. 3, 4.

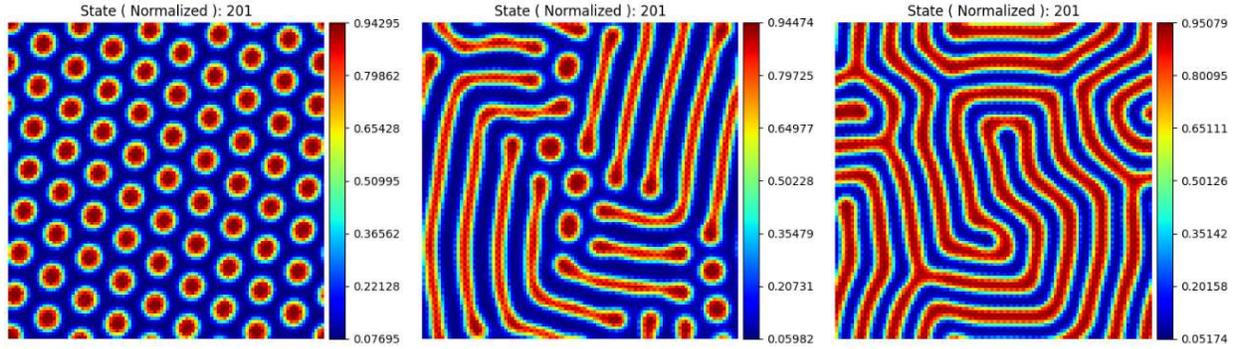

Fig. 3. Three main types of morphology in the case of decomposed steady-state: Gepard (spots), mixed morphology Gepard+Zebra, and Zebra (layered morphology – labyrinth or lamellae).

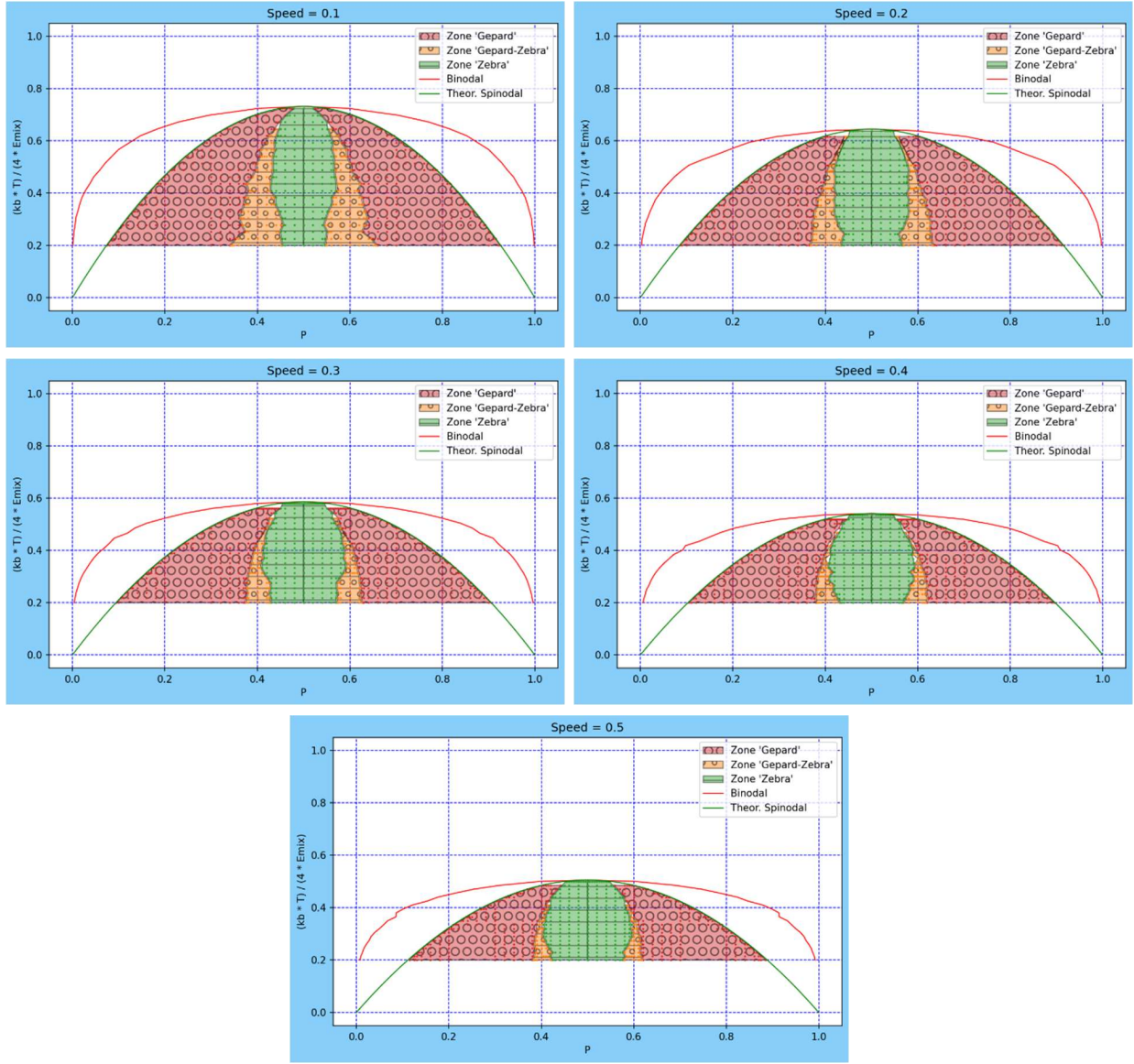

*Fig. 4. Map of steady-state morphologies, obtained with the initial condition of a homogeneous alloy with small noise amplitude under rates $v = 0.1, 0.2, 0.3, 0.4, 0.5$*

## 7. Influence of pre-existing structures: Decomposition beyond rate-dependent spinodal (as a result of composition or temperature shift from the spinodal region)

In the closed systems, ANY alloy between spinodal and binodal demonstrates decomposition by the nucleation-precipitation-coarsening mechanism. Since waiting for nucleation may last a very long time, one may use some pre-existing structures to initiate the decomposition. In our case of an open system, to accelerate the process, we use the structures formed at a previous composition or previous temperature. By this method, we managed to reach steady-state decomposition only within PART of the region between binodal and spinodal, attaching to the spinodal – see Fig. 5.

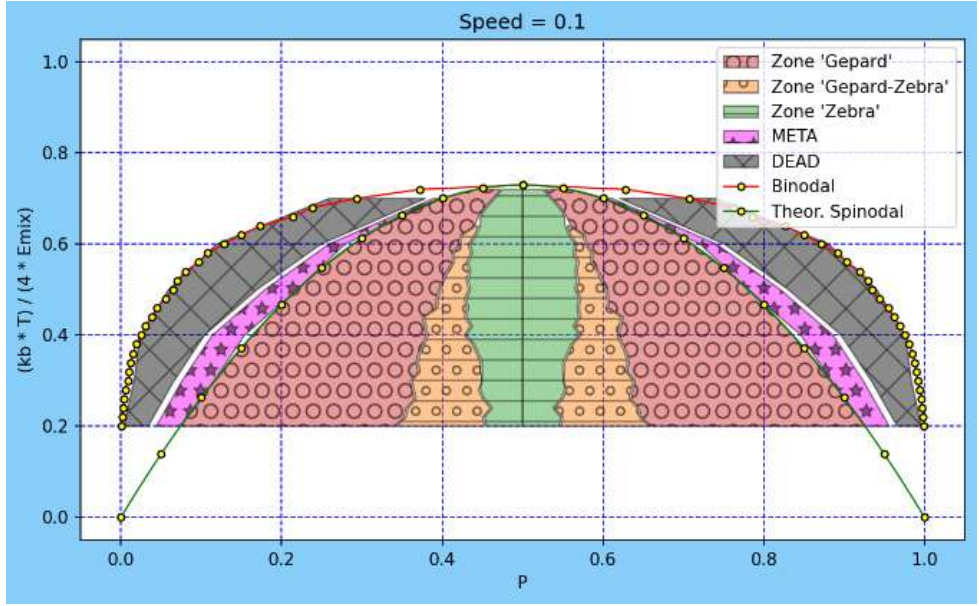

*Fig. 5. Violet regions beyond the spinodal demonstrating the formation of steady-state gepard-like morphology on the basis of pre-existing structures. Case $v = 0.1$*

## 8. Hysteresis.

As we can see from above, our model system demonstrates some memory of the final state regarding the initial conditions. This means that one may expect a hysteresis effect when each new morphology is obtained in the system using the previous morphology as the initial condition. At that, we may shift, step by step, the initial concentration (first from the right to the left, and then in the inverse direction – subsection 8.1), or we may shift, step by step, the temperature (first from low to high and then from high to low – subsection 8.2).

### 8.1. Compositional hysteresis.

At first, we obtain the steady-state pattern for the composition C=0.50 simulated from homogeneous initial conditions with a small initial noise amplitude of 0.001 at constant temperature kT/Emix=2. We take this as the initial condition for the sample with C=0.49, and so on, until C=0.20. Then we start to move back upwards in concentration, finishing with C=0.50. In Fig. 6, we compare the steady-state morphologies for C=0.50, obtained (first) from homogeneous initial conditions, and (second) as the last step of the step-by-step increase of C starting from 0.20. Two morphologies for C=0.49 are obtained from C=0.50 (and down) and from C=0.48 (and up), and so on. We can see a huge hysteresis with respect to the direction of the concentration shift: Going down from 0.50 preserves the Zebra morphology until C=0.38, and going up preserves the Gepard morphology up to C=0.47.

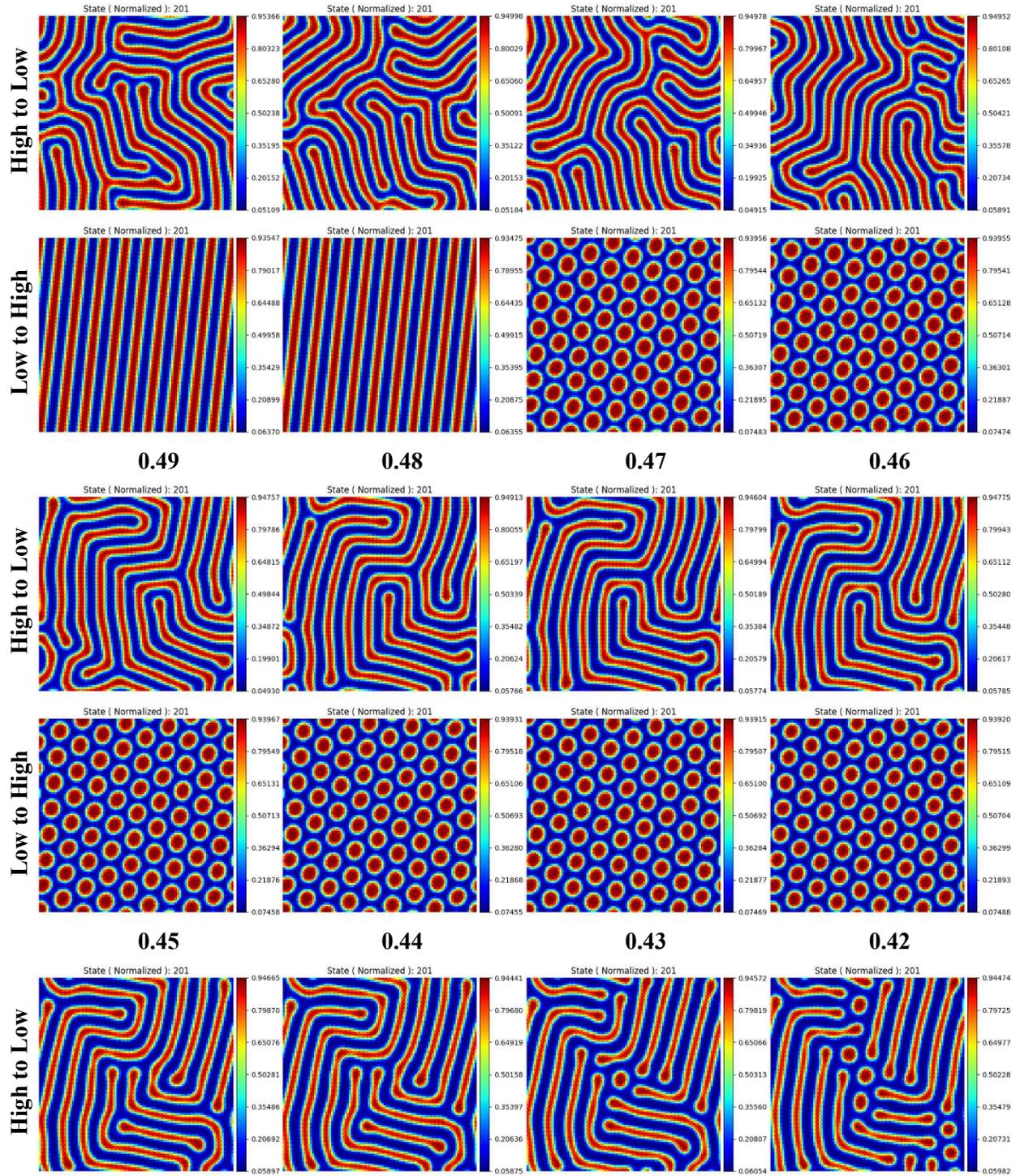

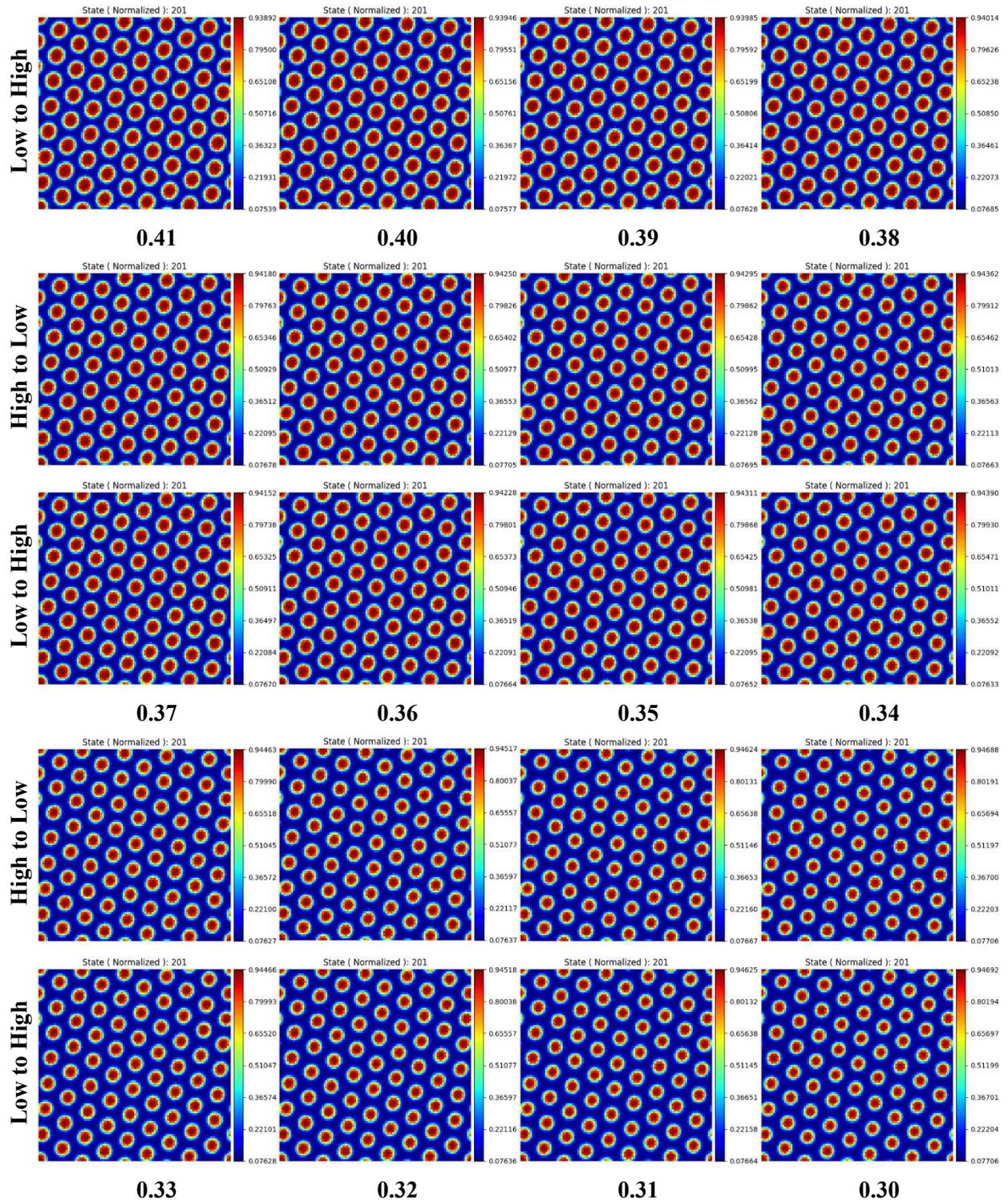

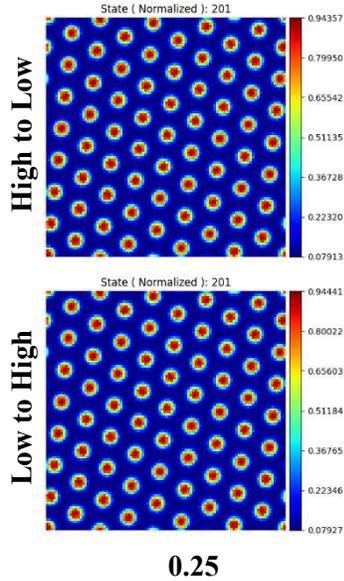

0.25

Fig. 6. Comparison of steady-state morphologies for the same composition, obtained from the preexisting structures with higher and lower concentrations of A as the initial condition.

### 8.2. Temperature hysteresis.

We took a constant concentration C=0.30 and changed the temperature kT/Emix from 0.8 to 2.4 and then back down. We can see that the type of morphology is almost the same (if one forgets about faceting), but the mean size and interparticle distances differ.

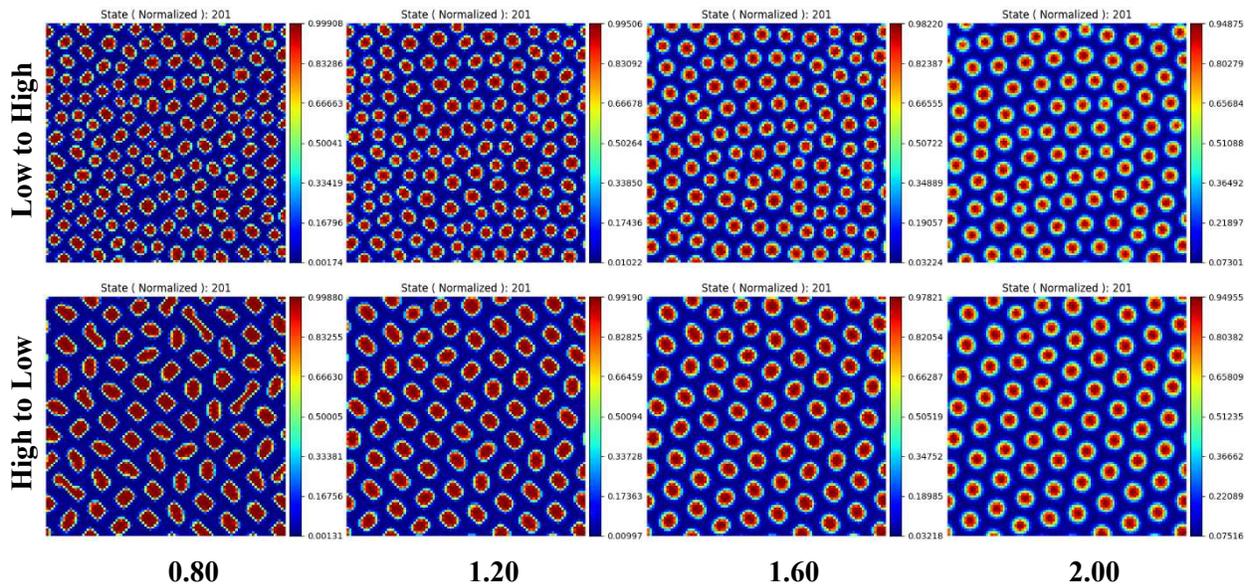

0.80      1.20      1.60      2.00

Fig. 7. Comparison of steady-state morphologies for the same temperature, obtained from the preexisting structures for higher and lower temperatures as the initial condition.

## 9. Evolution criterion.

For a closed system in a thermal bath at fixed temperature and volume, the evolution criterion should coincide with the second law of thermodynamics – the time derivative of the Helmholtz free energy should be negative and tending to zero, meaning an equilibrium state with minimal free energy. Let us find out the behavior of free energy in our simplified model system (and check if it reduces to the minimization of free energy in the case of zero rate). In the general case, the change rate of free energy should consist of the following terms: (1) in-flux of free energy corresponding to the homogeneous solid solution $\frac{V}{\delta}F\{C^{dep}\}$ in all sites, (2) out-flux of free energy with the actual redistributed concentration $-\frac{V}{\delta}F\{C(r)\}$, (3) change rate due to atomic exchanges within the system $[dF/dt]_{inner} = \sum_{(i,in)}^{NZ\|/2}(\mu_{AB}(i) - \mu_{AB}(in))\frac{dC^{i,in}}{dt}$, where $\frac{dC^{i,in}}{dt}$ is a partial change rate of site "i" occupancy by A (and simultaneously the partial change rate of site "in" occupancy by B due to exchange only between these two sites:

$$\frac{dC^{i,in}}{dt} = -C_A(i)C_B(in)\Gamma_{AB}(A(i) \leftrightarrow B(in)) + C_B(i)C_A(in)\Gamma_{AB}(A(in) \leftrightarrow B(i))$$

As shown in [8], $[dF/dt]_{inner}$ can be reorganized to the form

$[dF/dt]_{inner} = -\nu_0 exp\left(-\frac{E^S}{kT}\right) * \sum_{(i,in)}^{NZ\|\overline{2}}(\mu_{AB}(i) - \mu_{AB}(in))[exp(\mu_{AB}(i)/kT) - exp(\mu_{AB}(in)/kT))] * C_B(i)C_B(in) * exp\left(\frac{E_B(i)+E_B(in)}{kT}\right)$, which is similar to the expression in Boltzmann derivation of the H-theorem and is always negative. Thus,

$\frac{dF}{dt} = \frac{V}{\delta}[F\{C^{dep}\} - F] + [dF/dt]_{inner}$, $[dF/dt]_{inner} \leq 0$. This property can be reformulated as

$$\left[exp\left(-\frac{V}{\delta}t\right)\frac{d}{dt}exp\left(\frac{V}{\delta}t\right)\right](F - F\{C^{dep}\}) \leq 0 \qquad (22)$$

Eq. (22) is a generalized evolution criterion for our model of the open system.

## Conclusions

1. We developed a model of a 2D regular solid solution for an open system, characterized by an additional rate parameter—the deposition rate, V. For non-zero constant V, this model system reaches a steady state rather than equilibrium.
2. We analytically derived and numerically validated the rate-dependent spinodal for alloys that are entirely unstable with respect to any fluctuation. Increasing V stabilizes the alloy, lowering the spinodal curve as shown in equations (20) and (21).
3. If the initial system composition is nearly uniform with minor fluctuations, and the deposited composition-temperature point lies below the rate-dependent spinodal, the resulting state stabilizes into one of two primary morphologies (Figs. 3,4): "cheetah"

(gepard-like, spot-like) or "zebra" (layered, in the form of labyrinths or parallel lamellae). In a narrow transition zone, a mixture of these two morphologies emerges.
4. The rate-dependent binodal has been determined only numerically, by directly measuring the marginal compositions in the asymptotic steady states (Fig. 1). At that, the "interdome region" at the phase diagram (points beyond spinodal but within binodal) is subdivided into two subregions –see next conclusion.
5. In the before-mentioned intermediate regions between the rate-dependent binodals and rate-dependent spinodals, the structure of steady states depends significantly on pre-existing structures (memory effect). For a homogeneous initial alloy with slight initial noise, decomposition is suppressed everywhere beyond the rate-dependent spinodal. If the initial structure is prepared by shifting composition or temperature into an instability region and then returning it beyond the spinodal after partial decomposition, the alloy's behavior varies. Closer to the spinodal (but still beyond it), the steady state tends toward a cheetah-type (gepard-like) decomposition; in the remaining area between the spinodal and binodal, the alloy stays uniform.
6. Our system shows both compositional and temperature hysteresis: by using a previous structure as the initial condition for a new simulation with an altered deposition composition or temperature, we demonstrate that steady states in alloys with the same composition, but different initial conditions, can lead to markedly different morphologies.
7. We derived a generalized evolution criterion (22) for the free energy of our open system.
8. We also studied the properties of even more simplified model - when both exchanges and interaction energies are taken into account only within single top-plane. The results are qualitatively the same, except preferrable orientations for lamellar structure - they will be published elsewhere.

## ACKNOWLEDGMENTS

Authors are grateful for the "ENSEMBLE3-Center of Excellence for Nanophononics, Advanced Materials and Novel Crystal Growth-Based Technologies" project (GA No. MAB/2020/14), carried out under the International Research Agenda programs of the Foundation for Polish Science that are co-financed by the European Union under the European Regional Development Fund and the European Union Horizon 2020 Research and Innovation Program Teaming for Excellence (GA No. 857543) for supporting this work. The publication was created as part of the project of the Minister of Science and Higher Education "Support for the Activities of Centers of Excellence Established in Poland under the Horizon 2020 Program" under contract No. MEiN/2023/DIR/3797.
Authors are also grateful to Prof. King Ning Tu, to Dr. Ihor Radchenko and to Anastasiia Titova for fruitful discussions of phase and structural transitions in open systems.